\documentclass[journal,twoculomn]{IEEEtran}
\usepackage{cite}
\ifCLASSINFOpdf
 \usepackage[pdftex]{graphicx}

\else

  \usepackage[dvips]{graphicx}

\fi
\usepackage{bm}
\usepackage{epsfig}
\usepackage{amsmath,amssymb}

\usepackage{algorithmic}
\usepackage{relsize}
\usepackage{array}

\ifCLASSOPTIONcompsoc
 \usepackage[caption=false,font=normalsize,labelfont=sf,textfont=sf]{subfig}
\else
 \usepackage[caption=false,font=footnotesize]{subfig}
 \fi
\usepackage{amssymb}
\DeclareMathAlphabet\mathbfcal{OMS}{cmsy}{b}{n}
\usepackage{color}

\usepackage{amssymb}
\begin{document}

\title{A New Reconfigurable Antenna MIMO Architecture for mmWave Communication}

\author{
    \IEEEauthorblockN{Mojtaba Ahmadi Almasi\IEEEauthorrefmark{1}, Hani Mehrpouyan\IEEEauthorrefmark{1}, Vida Vakilian\IEEEauthorrefmark{7}, Nader Behdad\IEEEauthorrefmark{2}, Hamid Jafarkhani\IEEEauthorrefmark{3}}\\
    \IEEEauthorblockA{\IEEEauthorrefmark{1}{\small{Department of Electrical and Computer Engineering, Boise State University,
    \{mojtabaahmadialm,hanimehrpouyan\}@boisestate.edu}}}
    \IEEEauthorblockA{\IEEEauthorrefmark{7}{\small{Department of Electrical and Computer Engineering and Computer Science, California State University, \{vvakilian\}@csub.edu}}}\\
    \IEEEauthorblockA{\IEEEauthorrefmark{2}{\small{Department of Electrical and Computer Engineering, University of Wisconsin-Madison,
    \{behdad\}@wisc.edu}}}\\
    \IEEEauthorblockA{\IEEEauthorrefmark{3}{\small{Center for Pervasive Communications and Computing, University of California, Irvine,
    \{hamidj\}@uci.edu}}
\vspace{-0pt}
}
\thanks{This project is supported in part by the NSF ERAS grant award number 1642865,1642536.}
}


\markboth{}%
{Shell \MakeLowercase{\textit{et al.}}: Bare Demo of IEEEtran.cls for IEEE Journals}
\maketitle
\thispagestyle{empty}
\begin{abstract}
The large spectrum available in the millimeter-Wave (mmWave) band has emerged as a promising solution for meeting the huge capacity requirements of the 5th generation (5G) wireless networks. However, to fully harness the potential of mmWave communications, obstacles such as severe path loss, channel sparsity and hardware complexity should be overcome. In this paper, we introduce a generalized reconfigurable antenna multiple-input multiple-output (MIMO) architecture that takes advantage of lens-based reconfigurable antennas. The considered antennas can support multiple radiation patterns simultaneously by using a single RF chain. The degrees of freedom provided by the reconfigurable antennas are used to, first, combat channel sparsity in MIMO mmWave systems. Further, to suppress high path loss and shadowing at mmWave frequencies, we use a rate-one space-time block code. Our analysis and simulations show that the proposed reconfigurable MIMO architecture achieves full-diversity gain by using linear receivers and without requiring channel state information at the transmitter. Moreover, simulations show that the proposed architecture outperforms traditional MIMO transmission schemes in mmWave channel settings. 

\end{abstract}

\section{Introduction}
Millimeter-Wave (mmWave) technology operating in the $30$-$300$ GHz range is emerging as a promising solution for the 5th generation (5G) wireless communication systems by supporting a larger user base and high speed wireless links~\cite{r1}. The existence of a large communication bandwidth at mmWave frequencies will enable mmWave systems to support multi Gigabits/sec speeds 
However, significant path loss, channel sparsity and hardware limitations are major  obstacles for the deployment of mmWave systems.

In order to combat the severe path loss at mmWave frequencies, researchers have proposed to use large directional gains and line-of-sight (LoS) links~\cite{r2}.
Another approach that partially ignores the hardware costs, is based on the use of massive antenna arrays at the transmitter and receiver to mitigate the propagation issues at mmWave frequencies~{\cite{r3}}. Fortunately, the shorter wavelengths at mmWave frequencies allows for the deployment of large number of antennas. When considering such massive antenna structures, three beamforming architectures have been proposed for mmWave systems: a) \textit{digital}~\cite{r4}, b) \textit{analog}~\cite{r5,r14}, and c) \textit{hybrid}~\cite{r6}. The latter is significantly different from the beamforming approaches that had been proposed for multiple-input multiple-output (MIMO) systems intended for the sub-$6$ GHz band.

The digital approach provides great flexibility in shaping the transmitted beams. However, it requires one radio frequency (RF) chain per antenna. This results in significant cost and complexity in massive mmWave MIMO systems. Further, this network may result in significant delay due to large number of channel parameters that must be estimated~\cite{r13}. As an alternative, the analog method applies phase shifters to shape the output beam with only one RF chain for all the antennas~\cite{r14}. Although it is energy efficient and cost effective, the network can only provide a highly directional beam which in no way addresses shadowing and channel sparsity at mmWave frequencies~\cite{r13}. A promising beamforming approach for the mmWave MIMO architecture is based on a combination of analog and digital beamforming, i.e., hybrid beamforming. The hybrid architecture aims to use the benefits of digital and analog architectures. That is, the digital stage deals with the shadowing and channel sparsity by using several RF chains, while the analog stage provides directivity gain~\cite{r13}.

Three types of connected networks have been proposed for the hybrid architecture. The first network is called the fully-connected network because each RF chain is connected to all antennas via phase-shifters~\cite{r7}. While generating highly directional beams, this network suffers from a complicated beam selection network. In order to reduce the complexity, the concept of a sub-connected network has been introduced in~\cite{r6}. In this architecture, each RF chain is connected to a sub-group of antennas, which lowers complexity at the cost of reducing the directionality of the beams. The third methodology has a completely different structure, where several RF chains are connected to a lens antenna array via switches~\cite{r8,r9}. The transceiver antenna is able to generate a few orthogonal beams to achieve multiplexing gain for better utilization of the bandwidth, and to preserve low hardware complexity. 


Aside from the advantages of the hybrid architecture with a lens antenna array, high path loss and channel sparsity still remain big issues in MIMO mmWave systems. Recently, the idea of using a single reconfigurable antenna in a MIMO system instead of a massive antenna array was proposed in~\cite{r10,r25}. In this work, which is limited to only $2\times 2$ MIMO systems, two reconfigurable antennas are deployed at the transmitter in which each antenna steers several beams toward the receiver that is equipped with omni-directional antennas. Further, the recently designed lens antennas in~\cite{schoenlinner2002wide,r19} might represent better characteristics compared to that of composite right-left handed (CRLH) leaky-wave antennas (LWAs) that was considered in~\cite{r10,r25}. Reconfigurable antennas for use in MIMO architectures have also been considered in other works such as~\cite{RecAntMag}. In~\cite{r26} the diversity gain of traditional MIMO systems are improved through the application of both reconfigurable antennas at the receiver and space-time block codes (STBCs). Subsequently, the technique in~\cite{r26} is extended to a system with reconfigurable antennas at both the transmitter and receiver sides in~\cite{r27}. Later, a coding scheme was proposed for reconfigurable antenna MIMO systems over frequency-selective fading channels in~\cite{r28}. However, unlike the work in this paper, the main goal in~\cite{r26,r27,r28} is to use the degrees of freedom provided by reconfigurable antennas to enhance the diversity order of the system. Hence, the designs in~\cite{r26,r27,r28}, in no way address the propagation issues specific to mmWave MIMO systems, namely the channel sparsity.

In this paper, motivated by the $2\times 2$ reconfigurable MIMO system in {\cite{r10,r25}} and taking advantage of the antenna architecture in~\cite{r19}, we propose an $N_t \times N_r$ reconfigurable antenna MIMO architecture for mmWave communication. The reconfigurable antenna in~\cite{schoenlinner2002wide,r19} can support the simultaneous transmission of multiple radiation lopes via one RF chain. When this reconfigurable antenna is used in a MIMO setup, it can be used to establish a large number of paths between the transceiver antennas. The number of paths can conceivably overcome the channel sparsity issue. Subsequently, a rate-one STBC is used to encode the information symbols to achieve full-diversity gain, which can be applied to compensate for severe path loss at mmWave frequencies. Analytical results show that the proposed reconfigurable antenna MIMO architecture with STBCs achieves full-diversity gain by using linear receivers, e.g., zero-forcing (ZF). Further, simulations show that the proposed architecture outperforms traditional MIMO transmission schemes in mmWave channel settings. 

The paper is organized as follows: Section~II presents the reconfigurable antenna MIMO architecture and the system model. In Section~III, the signal processing algorithm for the proposed reconfigurable antenna MIMO architecture is derived and presented. In Section~IV, we present simulations investigating the performance of the proposed architecture, while comparing its bit error rate (BER) with that of traditional MIMO systems. Section V concludes the paper.

\textbf{Notations:} Hereafter, $j = \sqrt{-1}$, small letters, bold letters and bold capital letters will designate scalars, vectors, and matrices, respectively. Superscript $(\cdot)^{\dagger}$ denotes the transpose operator. vec$(\cdot)$ denotes the vectorization of $\mathbf{A}$ which is a column vector obtained by stacking the columns of the matrix on top of one another. Further, $\mathbb{E}(\cdot)$, $|\cdot|$, and $||\cdot||^2$ denote the expected value, absolute value, and norm-$2$ of $(\cdot),$ respectively. Finally, $\mathbf{A} \circ \mathbf{B}$ stands for the Hadamard product of matrices $\mathbf{A}$ and $\mathbf{B}$.

\section{Reconfigurable Antenna MIMO and System Model}
In this section, we present the advantages and disadvantages of hybrid beamforming systems, specifically for lens antenna arrays. Further, we outline the proposed reconfigurable antenna MIMO system and the corresponding system model.

\subsection{Overview of Lens Antenna Array System}
\begin{figure}
\includegraphics[scale = 0.82]{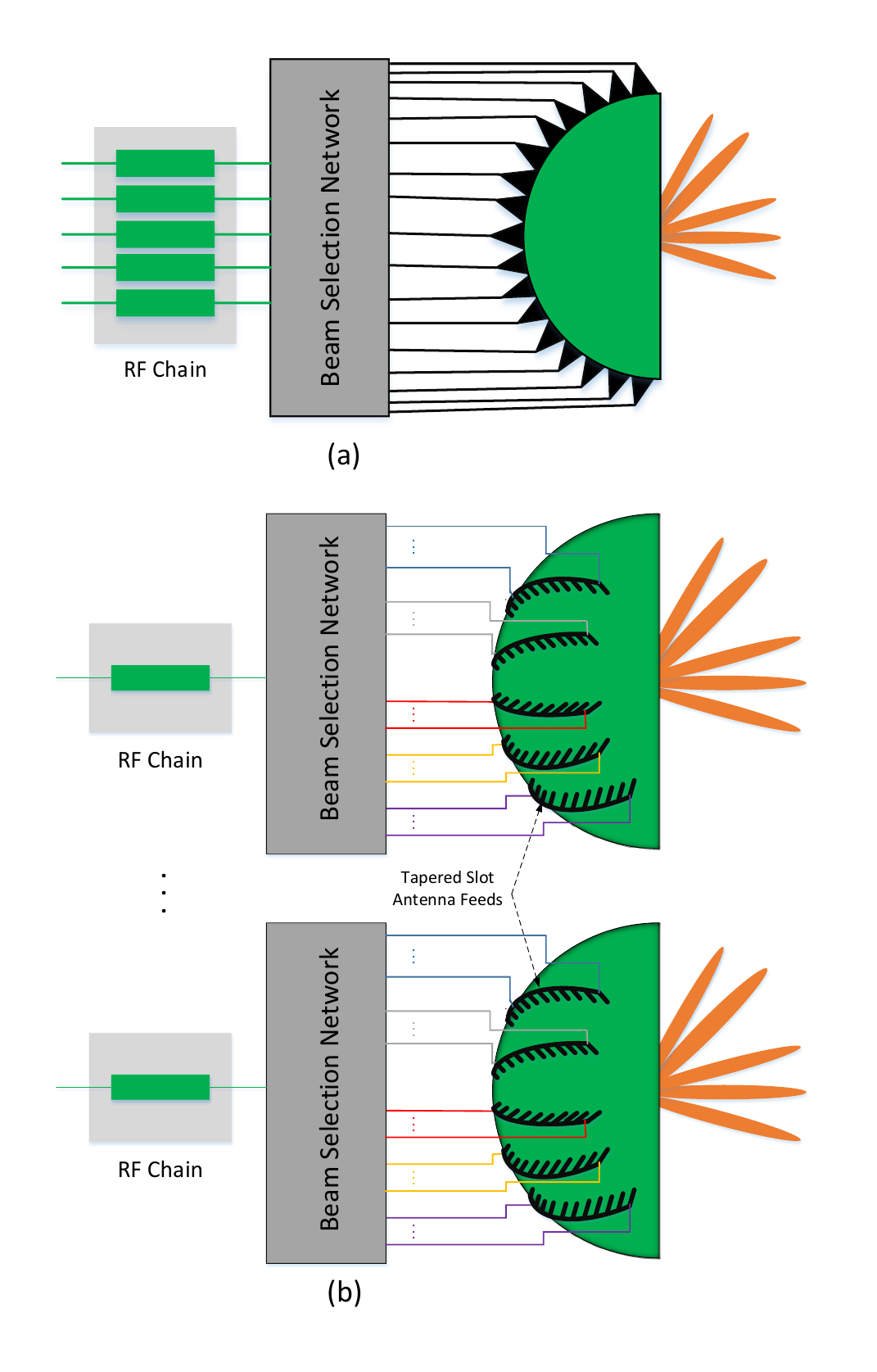}
\centering
\caption{Lens antennas with different architectures: (a) Beamspace MIMO transceiver \cite{r9}; (b) The proposed reconfigurable antenna MIMO transceiver based on lens antennas in \cite{r19}.}
 \label{fig0}
\end{figure}

 As stated before, the hybrid architecture provides a balance between the benefits of the digital and the analog architectures. Recall that three networks have been proposed for the hybrid architecture. In the first network, each RF chain is fully connected to all antennas through phase-shifters. 
 In the second network, each RF chain is connected to the sub-array of antennas through phase-shifters. 
 The third approach is based on lens antenna array networks. Unlike the prior two networks, this network uses a selection network that utilizes RF switches instead of phase-shifters~{\cite{r9}}. Further, the antenna structure is a feed antenna array placed beneath the lens which provides significant directivity gain. Thanks to the lens antenna, this structure generates narrow and high gain beams. The structure of the transceiver is shown in Fig.~\ref{fig0}.(a). 
 
 Let us focus on the third beamforming approach above. Since the mmWave environment does not result in rich scattering, the number of effective propagation paths, $p$, between the transceiver antennas is considerably less than the number of radiation beams, $n$, i.e., the number of antenna elements connected to the lens. Hence, the complexity of the selection network can be reduced as well as the number of RF chains without significant performance loss~{\cite{r9}}. Further, the authors of ~{\cite{r9}} show that this hybrid beamforming approach achieves a multiplexing gain of $p$ at mmWave frequencies. To summarize, this architecture, which is referred to as beamspace MIMO~\cite{r8,r9}, establishes communication paths between a single transceiver antenna pair through the multiple beams rather than multiple transceiver antenna pairs as shown in Fig.~\ref{fig0}.(a). Thus, the system is a beam-based MIMO rather than the traditional antenna-based MIMO systems. Throughout this paper, we refer to this system as the lens antenna array system. 
 
 
Although the methodology in~{\cite{r9}} is effective at mmWave frequencies, it may not be the most efficient approach. This follows from the fact that mmWave channels are sparse due to limited scattering and also suffer from dramatic path loss. Note that the lack of scattering can be attributed to significant path loss. This lack of scattering and the LoS nature of communication links, may result in ill-conditioned MIMO channels. In other words, the lens antenna array structure in~\cite{r8} and~\cite{r9}, fundamentally operates as a single input-output antenna system with multiple input-output beams in which the number of beams, $n$, is always limited to the number of paths, $p$, where $p$ $\ll n$~\cite{r9}. Hence, the approach in~\cite{r9}, in no way provides a solution to overcome channel sparsity at mmWave frequencies. On the other hand, in this paper we propose a true MIMO reconfigurable antenna structure that ensures we have as many paths, $p$, as the number of antennas.

\subsection{Reconfigurable Antenna MIMO Architecture}
As mentioned, the lens antenna array system in~\cite{r9} does not reflect the definition of the traditional MIMO system. The basic idea behind the traditional MIMO systems is to increase the independent paths in a communication link. Unfortunately the maximum number of the paths in the lens antenna array is equal to $p \ll n$ due to the sparsity in mmWave MIMO channels.
One promising solution for this problem is using lens antenna array in multiple input-output fashion. To make this idea more practical, the reconfigurable lens antenna design that is proposed in~\cite{schoenlinner2002wide,r19}, will be used. 

In this paper, by considering the benefits of the reconfigurable antenna in~\cite{schoenlinner2002wide,r19}, we propose a new MIMO architecture for mmWave communications. The architecture has two fundamental properties. First, each antenna is connected to one RF chain that can simultaneously radiate multiple beams as shown in Fig.~\ref{eq0}.~(b). Whereas, in  the lens antenna array in~\cite{r9}, each antenna is connected to $p$ RF chains. However, in comparison to~\cite{r9}, our architecture needs more beam selection networks and lens antennas. In addition, each antenna can independently change the phase of each radiation pattern. Second, unlike the lens antenna array system in~\cite{r9}, the transmitter and receiver have been equipped with $N_t$ and $N_r$ reconfigurable antennas, respectively. The proposed reconfigurable MIMO architecture provides the degrees of freedom that ensures that the number of MIMO channel paths is dictated by the number of antennas and not by the mmWave channel. Further, multiplexing gain is still achievable through the proposed design. Finally, we will show that merging the proposed reconfigurable antenna MIMO architecture with STBCs provides full-diversity gain that can be used to combat the significant path loss and shadowing at mmWave frequencies.

\subsection{System Model}
\begin{figure}
\includegraphics[scale = 0.34]{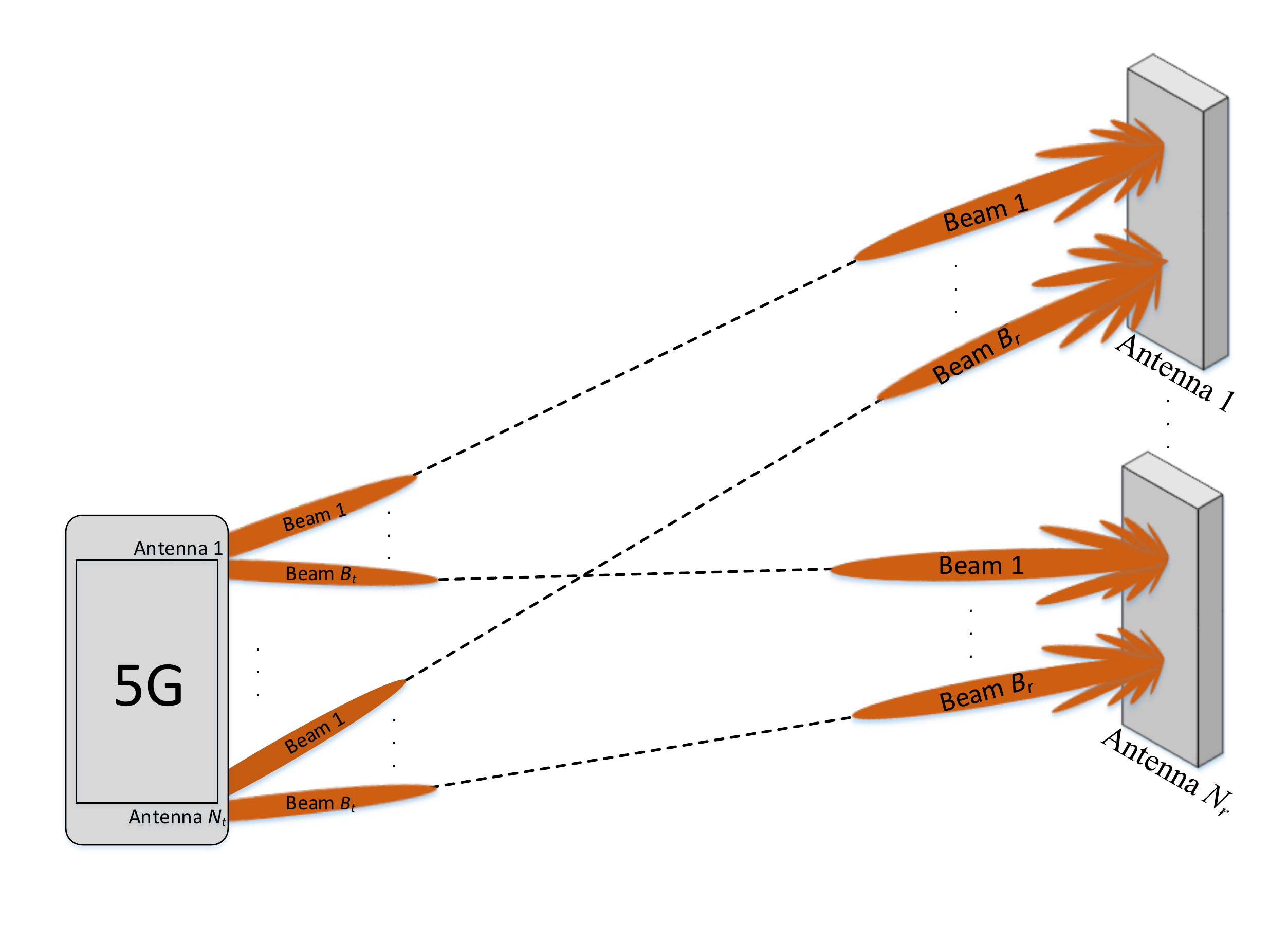}
\centering
\caption{An uplink MIMO system with reconfigurable antennas for 5G mmWave communications.}
 \label{fig1}
\end{figure}
Consider the uplink of a mmWave MIMO system that connects a single mobile user (MU) and base station (BS) (see Fig.~\ref{fig1}). It is assumed that the MU and BS are equipped with $N_t$ and $N_r$ reconfigurable antennas, respectively, where each transmit and receive antenna simultaneously radiate $B_t$ and $B_r$ beams, respectively. Moreover, the channel matrix between the MU and BS is denoted by $\mathbf{H} = [h_{n_t,n_r}]_{N_t\times N_r}$, where its entries are assumed to follow the Rician distribution. This is motivated by the presence of a LoS component in mmWave communication systems. Note that in previous mmWave MIMO systems, because of the large number of tightly-packed antennas, there is a high levels of antenna correlation. Hence, these works have adopted the Saleh-Valenzuela channel model. However, since we are using a small number of antennas, the Rician model would suffice~\cite{r7,r8,r9}. 


As outlined before and in~\cite{r19}, the proposed reconfigurable antenna can simultaneously change the phase of each radiation pattern. Mathematically, we represent this property with matrix $\mathbf{G}_t = [g^t_{n_t,b_t}]_{N_t\times B_t}$ when the reconfigurable antennas are used at the transmitter (MU) and $\mathbf{G}_r = [g^r_{b_r,n_r}]_{B_r\times N_r}$ when at the receiver (BS). These matrices are called the reconfigurable antenna parameter matrices. Thus, the entries of $\mathbf{G}_t$ and $\mathbf{G}_r$ are assumed to be
\begin{align} \label{eq33}
g_{n_t,b_t}^t &= e^{j\theta_{n_t,b_t}^t},~n_t = 1, 2, \dots, N_t,~b_t = 1, 2, \dots, B_t, \nonumber\\
g_{b_r,n_r}^r &= e^{j\theta_{b_r,n_r}^r},~b_r = 1, 2, \dots, B_r,~n_r = 1, 2, \dots, N_r,
\end{align}
respectively. The number of the radiation patterns (beams) of each transceiver antenna, i.e., $B_t$ and $B_r$, is assumed to be constrained to $N_r$ and $N_t$, respectively.\footnote{This assumption can be easily generalized to any $B_t$ and $B_r$ values.} Thus, we assume $B_t = N_r$ and $B_r = N_t$. For instance, the reconfigure parameter $g^t_{n_t,b_t}$ is related to the $b_t$th radiation pattern of $n_t$th antenna for $n_t$ = $1, 2, \dots,N_t$ and $b_t$ = $1, 2, \dots, N_r$ as shown in Fig. \ref{fig1}.

To realize the connection between the channel matrix and the reconfigurable parameter matrices, assume that a typical information signal, $s$, is sent from the $b_t$th beam of the $n_t$th antenna and received at the $b_r$th beam of the $n_r$th antenna. First, this signal is multiplied by $g^t_{n_t,b_t}$ when it is radiated through the $b_t$th beam of the $n_t$th antenna. The signal, then, is multiplied by the channel coefficient between the $n_t$th and the $n_r$th antenna, i.e., $s\times g^t_{n_t,b_t}h_{n_t,n_r}$. At the receiver, it is multiplied by $g^r_{b_r,n_r}$. That is, the received signal is equal to $s\times g^t_{n_t,b_t}h_{n_t,n_r}g^r_{b_r,n_r}$ without considering the noise term. This expression indicates that there is a one-to-one mapping among the entries of $\mathbf{G}_{t}$ and $\mathbf{H}$ at the transmitter side and $\mathbf{G}_r$ and $\mathbf{H} $ at the receiver side. Hence, the Hadamard product can nicely describe this mapping. That is, the reconfigurability brings about a new matrix denoted by the reconfigured channel matrix or $\mathbf{H}_g$, where
\begin{equation} \label{eq0}
\mathbf{H}_g = \mathbf{G}_t\circ \mathbf{H}\circ \mathbf{G}_r.
\end{equation}
$\mathbf{H}_g$ = $(\mathbf{h}_{g,1}, \mathbf{h}_{g,2}, \dots, \mathbf{h}_{g,N_r})$ of size $N_t \times N_r$ with entries $\mathbf{h}_{g,n_r}$ of size $N_t\times 1$. Here, $\mathbf{h}_{g,n_r}$ denotes the reconfigured channel between the MU and the $n_r$th receiver antenna at the BS.

We have defined the reconfigurable parameter and the reconfigured channel matrices. Now, we express the relationship between the transmit and receive antennas as
\begin{equation}\label{eq1}
\mathbf{Y} = \mathbf{X}(\textbf{s})\mathbf{P}\mathbf{H}_g + \mathbf{Z},
\end{equation}
where $\mathbf{Y} = (\mathbf{y}_1, \mathbf{y}_2,\dots,\mathbf{y}_{N_r})$ is the $T \times N_r$ received signal matrix. $T$ is the number of time slots in each block. $\mathbf{X}(\mathbf{s})$ is a $T\times N_t$ STBC matrix and $\mathbf{s} = (s_1, s_2, \dots,s_L)^\dagger$ is the $L\times 1$ information signal vector, where its elements are drawn from constellation $\mathcal{A}$. $\mathbf{P}$ is an $N_t\times N_t$ precoder matrix and $\mathbf{H}_g$ of size $N_t \times N_r$ is defined in (\ref{eq0}). The elements of the $T\times N_r$ noise matrix, $\mathbf{Z} = (\mathbf{z}_1, \mathbf{z}_2, \dots,\mathbf{z}_{N_r})$, are modeled by independent identically distributed (i.i.d) complex Gaussian random variables with mean 0 and variance $\sigma^2$, i.e., $z_{t,n_r} \sim$ $\mathcal{CN} = (0,\sigma^2)$ where $\sigma = \sqrt{\mathbb{E}\{|z_{t,n_r}|^2\}}$ for $t = 1, 2, \dots, T$ and $n_r = 1, 2, \dots, N_r$. Compared to the traditional MIMO systems, (\ref{eq1}) contains the reconfigurable channel matrix ($\mathbf{H}_g$) rather than the channel matrix ($\mathbf{H}$). The parameter $\mathbf{G}_t$, $\mathbf{G}_r$, $\mathbf{P}$, and $\mathbf{X}(\mathbf{s})$ in (\ref{eq1}) are flexible and should be designed.

It is worth noting that the proposed reconfigurable antenna MIMO architecture wisely exploits the capability of the reconfigurable antennas and traditional MIMO systems. The reconfigurable antennas aim to reconfigure the channel matrix which possibly results in a better channel condition. On the other hand, the MIMO structure attains to increase the dimension of the system (number of the paths) in order to deploy STBCs and precoders as well. A suitable STBC can achieve full-diversity gain, and consequently improve the BER performance of the system. On the other hand, precoders can properly allocate the power to further improve performance.

\section{Design of System Parameters}
In this section, we present the specific steps for selecting the appropriate values for $\mathbf{G}_t$, $\mathbf{G}_r$, $\mathbf{X}(\mathbf{s})$, and $\mathbf{P}$.

In a mmWave MIMO system, the channel state information (CSI) varies rapidly compared to the direction of each beam which helps to estimate the steering direction and CSI, separately~\cite{r11}. Since the steering direction changes slowly, this information can be useful at the transmitter. However, the fast varying CSI is of little value at the transmitter by the time it is estimated at the receiver and fed back. In our design we assume that there is no CSI available at the transmitter. However, the steering information is available at both the transmitter and receiver. In this case, similar to traditional MIMO systems, the optimum scenario is that the power is equally allocated at each antenna and beam, i.e., $\mathbf{P} = \mathbf{I}$. Further, the entries of $\mathbf{G}_t$ are considered to be one which yields
\begin{equation}\label{eq44}
\mathbf{H}_g =  \mathbf{H} \circ \mathbf{G}_r.
\end{equation}
Note that the use of the proposed reconfigurable antennas at the transmitter is motivated by the significant directionality gain that these antennas provide which is crucial at mmWave frequencies. Further, we will drop the subscribe $r$ of the $\mathbf{G}_r$ because only the receiver reconfigurable antennas will possibly effect the system. Hence, we have $\mathbf{H}_g =  \mathbf{H} \circ \mathbf{G}$
where the entries of $\mathbf{G}$ are equal to $e^{\theta_{b_r,n_r}}$. Thus, only two parameters $\theta_{b_r,n_r}$ and $\mathbf{X}(\mathbf{s})$ remain to be designed.

\subsection{Design of $\theta_{b_r,n_r}$}
While a signal is received through multiple paths, the best technique to increase the received signal-to-noise ratio (SNR) is to use maximum ratio combining (MRC)~\cite{r12}. Utilizing an MRC technique requires to weight the signals proportional to the inverse of the channel coefficient amplitudes before combining. Based on the proposed antenna design (these antennas can only change phase of each beam not the gain), this technique cannot be accomplished.

\begin{figure}
\includegraphics[scale = 0.5]{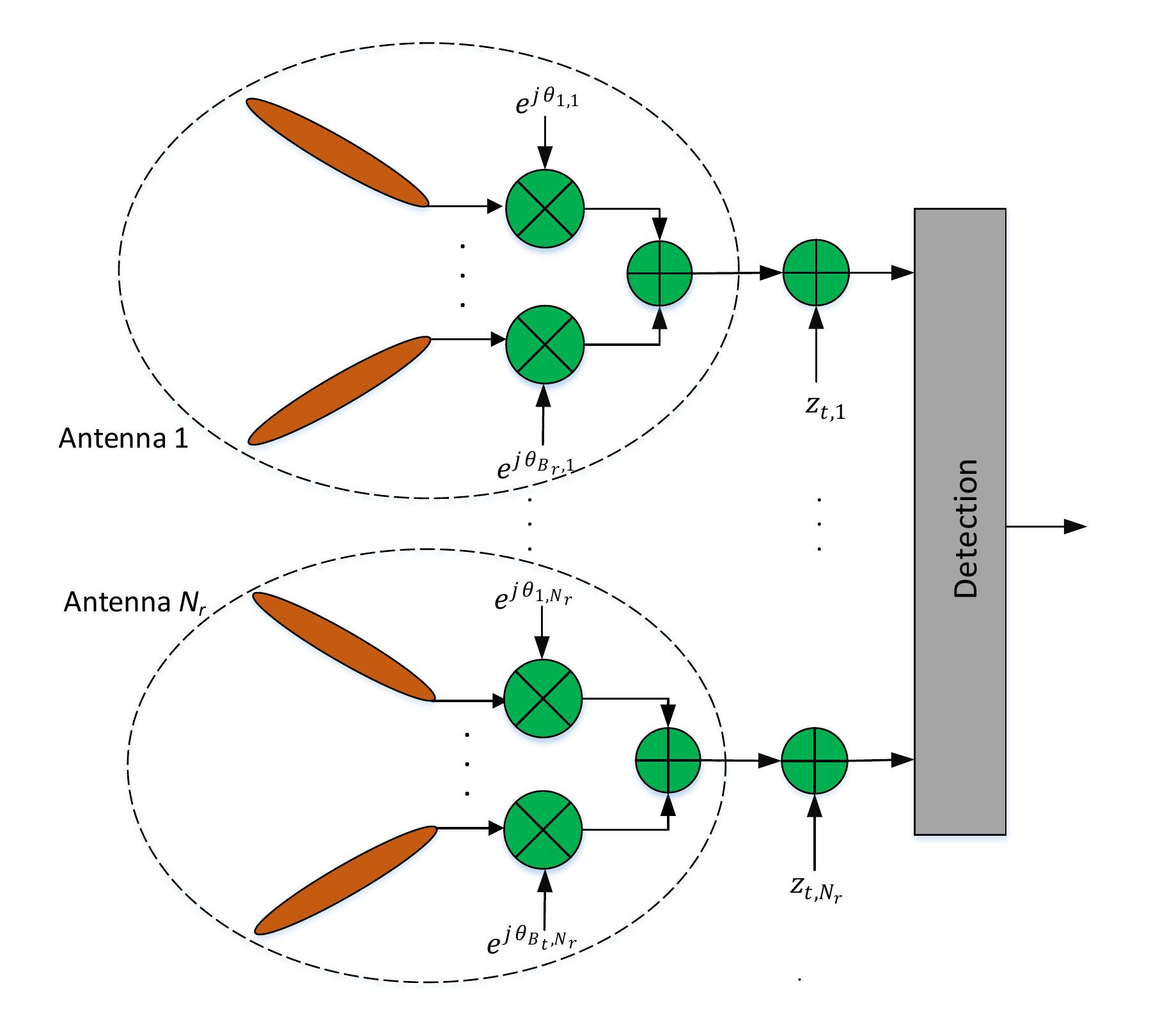}
\centering
\caption{The proposed receiver structure for the mmWave MIMO systems by regarding reconfigurable antennas.}
 \label{fig2}
\end{figure}

The alternative technique to improve the received SNR is the equal gain combining (EGC) {\cite{r12}}. It turns out that EGC is a suitable technique to design $\theta_{n_b,n_r}$s. This technique equally weights and co-phases the signals, which satisfies the proposed antenna design. Therefore, we have that
\begin{align} \label{eq66}
\theta_{b_r,n_r} = -\angle h_{n_t,n_r},~b_r, n_t = 1, 2, \dots,N_t, \nonumber \\
\quad n_r = 1, 2, \dots,N_r, \quad
\end{align}
where the sign $\angle$ stands for the phase of $h_{n_t,n_r}$. Thus,
$g_{b_r,n_r} = e^{-\angle h_{n_t,n_r}}$. Fig.~\ref{fig2} illustrates the receiver structure of our proposed MIMO system for a specific time slot $t$. The signal at each beam is multiplied by $g_{b_r,n_r}$. Obviously, (\ref{eq66}) leads to a non-negative real-valued matrix in (\ref{eq44}) while the channel matrix is complex-valued.
\subsection{Design of the STBC Matrix $\mathbf{X}(\mathbf{s})$}
One of our goals in this paper is to attain full-diversity gain in the proposed reconfigurable antenna MIMO structure. To this end, space-time block coding is a promising method. STBCs have been designed for both real and complex channels~\cite{HJbook}. Normally, real STBCs are used for transmission over real-valued channels and the complex STBCs are utilized for transmission over complex-values channels. Note that the channel matrix, $\mathbf{H}$, in our system is complex-valued and naturally a complex STBC can be used. However, we suggest compensating for the channel phase, as done in (\ref{eq66}) and the use of a real STBC. Since rate-one orthogonal STBCs (OSTBCs) that provide the maximum diversity and a simple symbol-by-symbol decoding exist for any number of transmit antennas, this will allow us to utilize a rate-one STBC and reduce the overall complexity by using linear receivers such as ZF. Note that rate-one complex OSTBCs do no exist for more than two antennas \cite{r15}.

A set of real-valued square STBCs for $N_t = 2, 4,$ and $8$ has been designed in~{\cite{r15}}. These STBCs are good candidates for our reconfigurable antenna MIMO even though here we use them with complex constellation points. Two examples of rate-one $\mathbf{X}(\mathbf{s})$ for $N_t = 2$ and $4$ are given by~\cite{r15}
\begin{equation} \label{eq77}
\mathbf{X}(\mathbf{s}) =
\begin{pmatrix}
s_1 & s_2 \\
-s_2 & s_1
\end{pmatrix}
\end{equation}
and
\begin{equation} \label{eq777}
\mathbf{X}(\mathbf{s}) =
\begin{pmatrix}
s_1 & s_2 & s_3 & s_4 \\
-s_2 & s_1 & -s_4 & s_3\\
-s_3 & s_4 & s_1 & -s_2\\
-s_4 & -s_3 & s_2 & s_1
\end{pmatrix},
\end{equation}
respectively. Note that while the codes are similar to the STBC in (\cite{r15}, Eq. (3) and (4)), the way we use them in the system, as outlined later, is different.  Other codes for any number of antennas can be found in {\cite{r15}}.
All these codes can be represented by  
\begin{equation} \label{eq20}
\mathbf{X}(\mathbf{s}) = \mathbf{A}_1s_1 + \mathbf{A}_2s_2 + \dots
+ \mathbf{A}_Ls_L.
\end{equation}
where $\mathbf{A}_i$s of size $T\times N_t$ are Hurwitz-Radon family of matrices which their entries are restricted to the set $\{-1,0,1\}$. Note that the information symbols $s_l$ for $l = 1, 2, \dots, L$ in (\ref{eq20}) are drawn from a complex-valued constellation whereas the STBCs in \cite{r15} encode real-valued symbols. Further, in \cite{r15}, it is proved that these matrices have the following properties:
\begin{align} \label{eq21}
\mathbf{A}_k^\dagger\mathbf{A}_k &= \mathbf{I}_{N_t}, \quad\quad\quad\quad k = 1,2, \dots, L, \nonumber \\
\mathbf{A}_k^\dagger\mathbf{A}_l &= -\mathbf{A}_l\mathbf{A}_k^\dagger,  \quad\quad 1 \leq k < l \leq L.
\end{align}
where $\mathbf{I}_{N_t}$ is the $N_t\times N_t$ identity matrix.
\subsubsection*{Diversity analysis} Designing a low-complexity decoding scheme for rate-one and full-diversity complex  STBCs is a big challenge. Among the available complex STBCs only the OSTBC in \cite{r16} achieves   rate-one, full-diversity gain and a simple symbol-by-symbol decoding. Unfortunately, for $N_t > 2$, the full-diversity complex STBCs cannot achieve rate-one for linear receivers \cite{r29}. However, we show that the STBC in (\ref{eq20}) attains full-diversity gain when linear receivers are considered.

By plunging (\ref{eq20}) into (\ref{eq1}) and after some manipulations, the received signal can be rewritten as
\begin{align} \label{eq22}
\bar{\mathbf{y}} &=  \big(\mathbf{b}_1, \mathbf{b}_2, \dots, \mathbf{b}_{L}\big)\mathbf{s} + \bar{\mathbf{z}},
\end{align}
where the $TN_r \times 1$ $\bar{\mathbf{y}} = \text{vec}(\mathbf{Y})$. Also, letting  $\text{vec}(\mathbf{H}_g) = \text{vec}(\mathbf{h}_{g,1}, \mathbf{h}_{g,2},\dots,\mathbf{h}_{g,N_r})$, $\mathbf{b}_{l}$ can be represented as $\mathbf{b}_{l} = \text{vec}(\mathbf{A}_l\mathbf{h}_{g,1}, \mathbf{A}_l\mathbf{h}_{g,2}, \dots,\mathbf{A}_l\mathbf{h}_{g,N_r})$ for $l = 1, 2, \dots, L$. Finally, $\bar{\mathbf{z}} = \text{vec}(\mathbf{Z})$. Defining $\boldsymbol{\mathcal{H}}_g$ as the reconfigured equivalent channel matrix, it can be expressed as
\begin{equation*}
\boldsymbol{\mathcal{H}}_g = \big(\mathbf{b}_1, \mathbf{b}_2, \dots, \mathbf{b}_{L}\big),
\end{equation*}
which is a real-valued matrix. This matrix exists when $\mathbf{H}_g$ is a non-zero matrix. Thus, (\ref{eq22}) can be rewritten as $\bar{\mathbf{y}} = \boldsymbol{\mathcal{H}}_g\mathbf{s} + \bar{\mathbf{z}}$. On the other hand, it is shown by \cite{r24} that for any $n \times 1$ vector $\mathbf{v}$ we have
\begin{equation} \label{eq23}
\mathbf{v}^\dagger\mathbf{A}_k^{\dagger} \mathbf{A}_l\mathbf{v} = \delta_{kl}\mathbf{v}^{\dagger} \mathbf{v},~k, l = 1,\dots,L,
\end{equation}
where $\delta_{k,l}$ is the Kronecker delta function and is 1 when $k = l$ and 0 when $k \neq l$. Regarding real-valued $\boldsymbol{\mathcal{H}}_g$ and referring to (\ref{eq23}), we can derive that 
\begin{equation} \label{eq24}
\boldsymbol{\mathcal{H}}_g^\dagger \boldsymbol{\mathcal{H}}_g= ||\mathbf{H}_g||^2\mathbf{I}_{L}.
\end{equation}
The above equation contains an important point. The reconfigured equivalent channel matrix is orthogonal which means the linear receivers will achieve full-diversity gain \cite{r29} for the STBC in (\ref{eq20}).

Considering a $2\times 1$ MIMO system, the reconfigured equivalent channel matrix of the code in (\ref{eq77}) with $\mathbf{h}_1 = (h_{1,1}, h_{2,1})^\dagger$, and equivalently $\mathbf{h}_{g,1} = (h^{g}_{1,1} = e^{j\theta_{1,1}}h_{1,1}, h_{2,1}^g = e^{j\theta_{2,1}}h_{2,1})^\dagger$ for the $\theta_{1,1}$ and $\theta_{2,1}$ defined in (\ref{eq66}), is given by
\begin{equation} \label{eq9}
\boldsymbol{\mathcal{H}}_g =
\begin{pmatrix}
h_{1,1}^g & \quad h_{2,1}^g \\
h_{2,1}^g & -h_{1,1}^g
\end{pmatrix}.
\end{equation}
Similarly, the reconfigured equivalent channel matrix of the STBC in (\ref{eq777}) for $N_r = 1$ is expressed as
\begin{equation} \label{eq7}
\boldsymbol{\mathcal{H}}_g =
\begin{pmatrix}
h_{1,1}^g &\quad h_{2,1}^g & \quad h_{3,1}^g & \quad h_{4,1}^g \\
h_{2,1}^g & -h_{1,1}^g & \quad h_{4,1}^g & -h_{3,1}^g \\
h_{3,1}^g & -h_{4,1}^g & -h_{1,1}^g & \quad h_{2,1}^g \\
h_{4,1}^g & \quad h_{3,1}^g & -h_{2,1}^g & -h_{1,1}^g \\
\end{pmatrix}.
\end{equation}
The above matrices are real-valued and obviously orthogonal.
\subsection*{The following remarks are in order:}
\begin{itemize}
\item{As mentioned, there is a specific difference between the SBTCs in {\cite{r15}} and the ones in this paper. The STBC defined by (\ref{eq20}) encodes complex-valued symbols whereas {\cite{r15}} encodes real-valued symbols. Nevertheless, in both cases, the code rate is equal to one and the codes facilitate low complexity decoding.}
\item{
To encode complex-valued information symbols for $N_t = 2$ and $4$, the OSTBC {\cite{r16}} and the Quasi-OSTBC (QOSTBC) {\cite{r17}} are well-suited rate-one codes. However, those codes are designed for a complex-valued channel matrix in which $s$ and conjugate of $s$ are encoded to achieve full-diversity gain. Although the channel matrix in our systems is complex-valued, thanks to the reconfigurable antennas and (\ref{eq66}) the reconfigured channel matrix becomes real-valued. Therefore, the conjugate of the information symbols is not required.}
\item{While real OSTBCs exist for any number of antennas, for more than four antennas, low complexity rate-one full-diversity complex STBCs are not available. As such, in addition to low complexity and full diversity, the rate advantage of the proposed system will result in a big coding gain for large number of transmit antennas.
}
\end{itemize}

\section{Numerical Results}
This section discusses the results of the numerical simulation for the proposed structure in Fig.~\ref{fig2} along with the STBCs in (\ref{eq77}) and (\ref{eq777}) and ZF receiver. Moreover, for traditional MIMO systems, the OSTBC and QOSTBC with maximum likelihood (ML) decoding are considered. Further, it is assumed that the lens array antenna system in~\cite{r8,r9} transmits one independent stream of symbols per RF chain to achieve multiplexing gain when using a ZF receiver. The communication channel is modeled as a Rician fading channel, i.e.,
\begin{align}
\mathbf{H} &= \sqrt{\frac{K}{K+1}}\mathbf{H}_L + \sqrt{\frac{1}{K+1}}\mathbf{H}_\textit{NL},
\end{align}
where $K$ is the Rician $K$-factor expressing as the ratio of powers of the LoS signal and the scattered waves. Throughout the simulation, $K$ is set to $2$ dB. Using this model, $\mathbf{H}$ is decomposed into the sum of the deterministic component $\mathbf{H}_{\textit{L}}$ and a random component $\mathbf{H}_{\textit{NL}}$. The former models the LoS signals. In the simulation, the entries of matrix $\mathbf{H}_L$ are all set to one. This is since the optimal LoS MIMO channels are highly dependent on the distance between the transmitter and the receiver, and the antenna spacing~\cite{r23}. These conditions cannot be easily satisfied in mobile cellular networks. Hence, here, an ill-conditioned LoS channel has been considered. $\mathbf{H}_{\textit{NL}}$ accounts for the scattered signals with its entries being modeled as i.i.d with $\mathcal{CN} \sim (0,1)$. Further, it is assumed that for OSTBC and QOSTBC the transceiver antennas are the traditional omni-directional antennas and for (\ref{eq77}) and (\ref{eq777}) they are the reconfigurable antennas as shown in Fig.~\ref{fig0}. (b). 
\begin{figure}
\includegraphics[scale = 0.83]{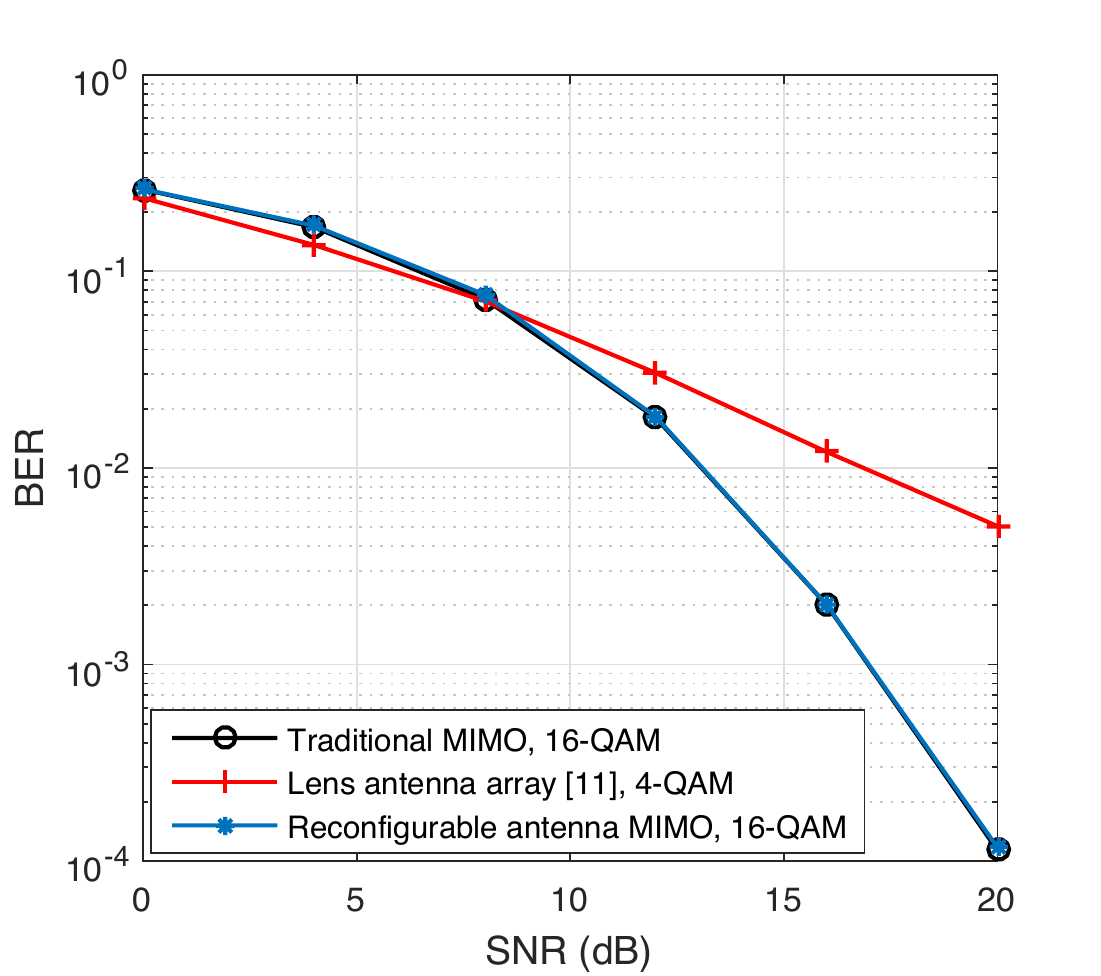}
\centering
\caption{BER performance of a $2\times 2$ traditional MIMO, lens array antenna and the proposed reconfigurable antenna MIMO.}
 \label{fig3}
\end{figure}

\begin{figure}
\includegraphics[scale = 0.82]{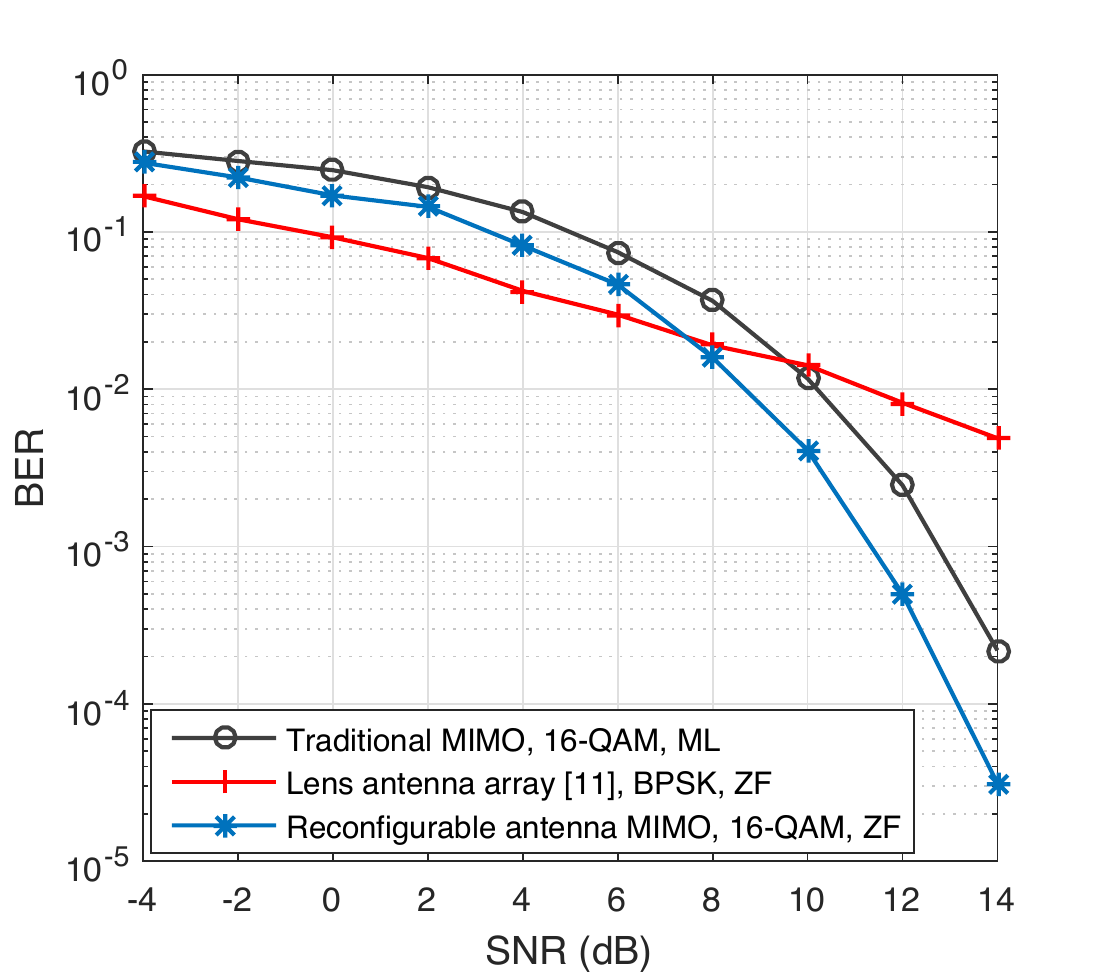}
\centering
\caption{BER performance of QOSTBC for $4\times 4$ reconfigurable antenna MIMO and the traditional MIMO systems.}
 \label{fig4}
\end{figure}

Fig.~\ref{fig3} shows the BER performance of (\ref{eq77}), OSTBC in \cite{r16} and the lens array antenna~\cite{r8,r9} versus SNR for a throughput of $4$ bits/s/Hz. The symbols for the traditional and the reconfigurable antenna MIMO are drawn from $16$-QAM and for the lens array are drawn from 4-QAM. It is assumed that the $N_t = 2$ and $N_r = 2$. The results show that (\ref{eq77}) and the OSTBC attain identical performance with full-diversity gain. Whereas, the lens antenna array shows the diversity gain of one since only there is one path between each transceiver pair. At low SNR the array slightly outperforms, whereas by increasing the SNR its performance decreases. This follows since the STBC in (\ref{eq77}) and OSTBCs are designed to achieve full-diversity gain at high SNRs. We use a ZF receiver for all simulations in this case. 


Fig.~\ref{fig4} compares the BER performance of (\ref{eq777}), the QOSTBC in {\cite{r17}} and the lens array antenna for a $4 \times 4$ MIMO system. Similar to the previous scenario the throughput is considered as $4$ bits/s/Hz. The constellation size for the reconfigurable antenna MIMO is set to $16$-QAM and for the lens antenna array is set to binary phase shift keying (BPSK). For the QOSTBC, half of symbols are chosen from $16$-QAM and other half are chosen from rotated $16$-QAM, where the rotation value is set to $\phi = \frac{\pi}{4}$. Further pair-wise ML decoding is used for QOSTBC. Similarly, while assuming the same throughput, the lens antenna array has the lowest BER at SNRs of less than $7$ dB, but at high SNRs the performance gap between the lens antenna array approach and the proposed method widens since the lens antenna array does not achieve the same diversity order as the proposed system. Further, the result indicates that both (\ref{eq777}) and the QOSTBC achieve full-diversity gain. However, our proposed MIMO outperforms QOSTBC by $1.6$ dB. The reason for this performance is due to the orthogonality of the reconfigured equivalent channel matrix in (\ref{eq7}).

\section{Conclusion}
In this paper, we discuss the properties of lens array antenna MIMO. Then, we propose a new reconfigurable antenna MIMO system for mmWave wireless networks. The reconfigurable antennas are inspired from the lens antennas. This MIMO system is proposed to achieve two goals. The first goal is to combat the sparsity of the MIMO channel at mmWave frequencies. The proposed reconfigurable antenna structure provides the system with additional degrees of freedom that can be used to overcome channel sparsity at mmWave frequencies. The second goal is to suppress the effect of path loss in mmWave systems. To achieve this goal, we covert the complex-valued channel matrix to the real-valued one using the reconfigurable antennas. Then, by applying STBCs, full-diversity gain is attained. Simulations verify that the proposed reconfigurable antenna architecture achieves a rate of one while at the same time reaching full diversity gain. 
\appendices

\bibliographystyle{IEEEtran}
\bibliography{IEEEabrv,references}
\end{document}